\documentclass{bioinfo}
\usepackage{array}
\usepackage{float}
\copyrightyear{2017}
\pubyear{2017}

\begin{document}
\firstpage{1}

\title[SANA: Separating search from objective, Part 1]{SANA: separating the search algorithm from the objective function in biological network alignment, Part 1: Search}
\author[Kanne, Hayes]{Dillon Kanne, Wayne B. Hayes\footnote{to whom correspondence should be addressed ({\tt whayes@uci.edu})}\;}
\address{Department of Computer Science, University of California, Irvine CA 92697-3435, USA}

\history{Received on XXXXX; revised on XXXXX; accepted on XXXXX}

\editor{Associate Editor: XXXXXXX}

\maketitle


\begin{abstract}
Biological network alignment is currently in a state of disarray, with more than two dozen network alignment tools having been introduced in the past decade, with no clear winner, and other new tools being published almost quarterly. Part of the problem is that almost every new tool proposes both a new objective function and a new search algorithm to optimize said objective. These two aspects of alignment are orthogonal, and confounding them makes it difficult to evaluate them separately. A more systematic approach is needed. To this end, we bring these two orthogonal issues into sharp focus in two companion papers. In Part 1 (this paper) we show that simulated annealing, as implemented by SANA, far outperforms all other existing search algorithms across a wide range of objectives. Part 2 (our companion paper) then uses SANA to compare over a dozen objectives in terms of the biology they recover, demonstrating that some objective functions recover useful biological information while others do not. We propose that further work should focus on improving objective functions, with SANA the obvious choice as the search algorithm.

\textbf{Availability:} SANA is on github.

\textbf{Contact:} whayes@uci.edu

\textbf{Supplementary information:} None yet.
\end{abstract}

\section{Introduction}
Network alignment aims to find relationships between different networks or subsections of them. A common use of network alignment is comparing protein-protein interaction (PPI) networks to draw conclusions about function and topology-function relationships of proteins between species \citep{graal,topo-function}. Network alignment is a fundamentally difficult problem; it is a generalization of an already NP-Complete problem, subgraph isomorphism \citep{cook1971complexity}, and current data sets are very noisy \citep{falsepositives}. Therefore, modern alignment algorithms try to approximate solutions using heuristic approaches. These algorithms rely on two explicit, orthogonal components: a search algorithm to look through the incredibly large\footnote{The number of alignments is \(\frac{N_2!}{(N_2-N_1)!}\), where $N_2$ and $N_1$ are the number of nodes in the larger and smaller networks, respectively.} space of possible alignments and an objective function to score the quality of alignments and guide the search algorithm. PPI network alignment has been shown to transfer functional information between species \citep{graal} and individual proteins \citep{topo-function}, but nonetheless PPI network alignment remains in its infancy because frequently new search algorithms cannot be directly compared to previous ones; there is no gold standard to compare against. Many new algorithms boast a new search algorithm and/or a new objective function, but fail to separate the two, making proper comparison nearly impossible due to confounding variables. 

In the past few years pairwise biological network alignment algorithms have improved considerably, incorporating new types of search algorithms relying on greedy \citep{hubalign,proper,wave}, Lagrangian \citep{natalie,lgraal}, or random search methods \citep{sana1,optnetalign,magna++} and different methods of evaluating alignments such as Importance \citep{hubalign}, graphlet degree vector similarity \citep{graal}, EC \citep{graal}\footnote{EC is variously called edge correctness, coverage, or conservation by various authors, and measures the fraction of edges in the smaller network that align to edges in the larger one.}, Weighted EC \citep{wave,lgraal}. Most of the time, however, new alignment algorithms are published in tandem with new object functions, making it nearly impossible to disambiguate and compare the quality of alignments across different alignment algorithms. Although some topological standards have been used as a basis for comparison, like EC and $S^3$ (Symmetric Substructure Score; Saraph and Milenkovíc, 2014), new objective functions and search algorithms continue to be implemented in tandem \citep[and others.]{tame,great,proper,hubalign,ghost}. Other algorithms' publications even mention how the performance of objective functions has ``always been unclear'' \citep{optnetalign}. We attempt to disambiguate this problem in this paper and the companion paper \cite{companion}.

These problems remain at the forefront of the field, with recent publications continuing to describe how current network aligners' and their objective functions cannot be properly compared due to their differences \citep{ulign}. Because of this, they argue, there is ``no consensus\ldots on which aligners should be used\ldots and which [objective functions] should be used to evaluate them,'' and certainly no ``gold standard'' search algorithm or objective function for network alignment. We attempt to solve the problem of which network alignment algorithm to use in this paper and what objective function to optimize in the companion paper.

In this work we test many of these recent alignment algorithms and objective functions against our state-of-the-art network alignment algorithm, \textit{Simulated Annealing Network Aligner} (SANA). In the first SANA publication \citep{sana1} we showed that many search algorithms cannot recover the perfect alignment even when given the perfect objective function as input; SANA was the only algorithm that universally recovered the perfect alignment in seconds, in all cases tested \citep[Figure 2]{sana1}. This suggests that SANA is a superior search algorithm in the case that we are given the perfect objective function as input.

In this paper, we extend that result to show that SANA appears almost universally capable of optimizing any objective function better than the aligners that proposed those objective functions. By fairly comparing modern network aligners against SANA using their own objective functions, we show that SANA should be treated as the ``gold standard'' for network alignment that some authors do not believe currently exists \citep{ulign}. Our goal is to convince the reader that further work on search algorithms is pointless; just use SANA. The only remaining closely related question is then a fair evaluation of said objective functions, which is the subject of our companion paper \cite{companion}. 

It should be understood that this paper does not directly involve biological information or biological analysis of these objective functions. This subject is pursued in great depth in the companion paper, but the current work is necessary to show that SANA is capable to optimize the other objective functions better than anyone else and compare them fairly in that paper.

\section{Materials and Methods}

\subsection{Compared Algorithms and Objectives}
\label{subsec:comparedalgs}
We compared SANA against MAGNA++ \citep{magna++}, GHOST \citep{ghost}, GREAT \citep{great}, HubAlign \citep{hubalign}, NATALIE 2.0 \citep{natalie}, PROPER \citep{proper}, OptNetAlign \citep{optnetalign}, L-GRAAL \citep{lgraal}, WAVE \citep{wave}, and ModuleAlign \citep{modulealign}. Many of these algorithms cannot directly optimize global topological measures such as EC or $S^3$, but instead use those measures after aligning to evaluate their alignments. Instead, they usually optimize some other topology-related objective function, while also introducing a new search algorithm that attempts to optimize said objective function. These other objective functions are listed with brief descriptions in Table \ref{tab:aligners}. Although GRAAL \citep{graal} was the first algorithm to introduce graphlet degree vector similarity and EC, we do not compare against it since it is no longer competitive against algorithms that have come after it.

\begin{table*}
\begin{center}
\caption{Objective Functions and Alignment Algorithms Compared}
\label{tab:aligners}
\begin{tabular}{|| p{3.3cm} p{2cm} p{3cm} p{7.8cm} ||} 
 \hline
 Objective Function & Proposed By & Used By & Brief Description  \\ [0.5ex] 
 \hline\hline
 GDV Signature Sim. & GRAAL & MAGNA++, SANA & Local measure score based on similarity of surrounding substructures. \\ 
 \hline
 L-GRAAL GDV Sim. & L-GRAAL & L-GRAAL, SANA & L-GRAAL's modified GDV similarity, normalized differently and using only graphlets with 4 or fewer nodes.\\
 \hline
 Spectral Signature Sim. & GHOST & GHOST, SANA & Based on the distribution of eigenvectors of normalized Laplacian matrices of surrounding subgraphs of each node.\\
 \hline
 Edge Graphlet Vector Sim. & GREAT & GREAT, SANA & Similar to Graphlet similarity, but scores pairs of edges instead of pairs of nodes. \\
 \hline
 Importance & HubAlign & HubAlign, ModuleAlign, SANA & Ranks nodes recursively based on their degree and the importance of adjacent nodes. \\
 \hline
 NATALIE 2.0's Objective Function (Unnamed) & NATALIE 2.0 & NATALIE 2.0, SANA & Based on sequence and updated by Lagrangian relaxation. We chose to optimize NATALIE's final (best) similarity matrix.\\ 
 \hline
 Greedy EC + Degree Diff. & PROPER & PROPER, SANA & Greedily built ``local EC,'' ties broken by degree difference.\\
 \hline
 GDV-Weighted EC (WEC) & WAVE & WAVE, SANA & Overlapping edges receive a score based on the average of the graphlet degree difference scores of their endpoints.\\
 \hline
 EC & GRAAL & MAGNA++, OptNetAlign, SANA & The percent of edges in the smaller network that align to edges in the bigger one.\\ 
 \hline
 Symmetric Substructure Score ($S^3$) & MAGNA++ & MAGNA++, OptNetAlign, SANA & Similar to EC, but symmetric and measured only on the subgraph induced by the node alignment. \\
 \hline
\end{tabular}
\end{center}
\end{table*}

To explain our exact methodology we will expand on the objective function, search algorithm, and comparison details for each of the algorithms. We do not elaborate on the details of any of the objective functions listed in Table \ref{tab:aligners}; refer to the respective publications for those details.

\subsubsection{Random Search Algorithms}  \hfill
\label{subsubsec:randomsearch}

Random search algorithms generally start with a random alignment or set of alignments.  They then generate a new set of alignments using some random perturbation method, score the new set, and choose ``good'' ones to move forward.  The process stops when the alignment has ``converged,'' which simply means a local maximum (or maxima) has been reached and the alignment stops by decree.

{\bf SANA}, the present work, optimizes any objective function using simulated annealing to make many, fast incremental improvements while accepting some bad and some good moves to avoid local maxima. Those interested in a full understanding of our algorithm should see \citep{sana1} for details. SANA can accept any objective function as long as it provides enough guidance to determine whether an individual move is good or bad. It is preferable that the objective function can also be quickly incrementally evaluated; recomputing the score after a small change should be fast. We have either implemented the objective functions of other algorithms into the code of SANA or edited the other algorithms' code to produce a similarity matrix file which SANA can interpret. A similarity matrix file is a \textit{local} measure, which means that it gives node-to-node similarity values for each pair of nodes $(u_1,v_1)$, where $u_1$ and $v_1$ are from different graphs. It can be fully described using a node-to-node matrix of size \(N_2\times N_1\), where $N_2$ and $N_1$ are the number of nodes in each graph. The canonical example of a local measure is sequence similarity, but topological measures can also be local. 

SANA has the ability to optimize any weighting of any number of objective functions. In most cases comparing with other algorithms, we simultaneously optimized some amount of EC, $S^3$, and the other objective function, with the weights chosen judiciously so that for each pair of networks we produced an alignment that scored better than the other aligner at all three measures simultaneously. These weights can be seen in Table \ref{tab:balancing}. We emphasize that our main goal was to show that we could outperform each of the other search algorithms at optimizing their own objective; the addition of EC and $S^3$ to SANA's objective was simply to show that even if forced to balance topology with another objective, SANA can {\em still} beat the other search algorithm at its own objective function while simultaneously beating it at both EC and $S^3$, which are commonly used topological quality objective functions.

We gave SANA 20 minutes of annealing time for all tests. This means that SANA spent 20 minutes actually aligning the networks, but running SANA for a new pair of networks and a new combination of objective functions requires a few minutes of overhead to generate the temperature schedule (see Section \ref{subsec:temperature} for details). This means that most 20 minute runs of SANA took closer to 23 to 25 minutes to run. The temperature information is then cached so that the same execution in the future does not need the same overhead. This is much faster than most of the other algorithms, many of which take multiple hours or days to complete. The full chart of SANA's runtimes versus these other algorithms can be seen in the Supplementary Material. As a more stringent test we also performed the exact same sets of experiments for even shorter runs of SANA (3 and 8 minutes, respectively).

{\bf MAGNA++} is a genetic algorithm that evolves a population of ``good'' alignments by culling bad ones and creating a new generation based on the best ones of the previous generation. Like SANA, it can optimize any objective function. MAGNA \citep{magna} was the first paper to introduce $S^3$ (Symmetric Substructure Score) which has been widely regarded as a good topological objective.  We use MAGNA++'s default parameters and chose a 60-40 balance between $S^3$ and graphlet degree vector similarity for MAGNA, since the the authors \citep[Figure 1]{magna++} say this combination provides the best results.

{\bf OptNetAlign} is a memetic search algorithm that maintains a population of alignments along a Pareto front in an attempt to automatically balance several competing objectives. Like MAGNA++, OptNetAlign defaults to using $S^3$ as its topological measure, and this is how we compare against them. They note in their publication that OptNetAlign could be programmed to optimize other objective functions, but their current implementation only includes $S^3$.

\subsubsection{Seed and Extend Algorithms} \hfill

A {\em seed} is a pair of nodes $(u_1,v_1)$ from $G_1$ and $G_2$ respectively, initially aligned together typically because they are the most similar pair, by some measure, among currently unaligned nodes.  Once a seed pair (or pairs) is chosen, the alignment is {\em extended} outwards by looking for a pair of nodes adjacent to the currently aligned set that are also highly ranked. By their very nature, seed-and-extend algorithms impose some sort of topological connectivity on the alignment because the {\it extend} part must follow {\em edges} in both graphs outwards from the currently aligned set of nodes. However, the topology thus implicitly imposed is rarely studied (other than to compute the EC and $S^3$ after-the-fact) and may be ill-defined.

{\bf HubAlign} ranks nodes by a measure called {\em Importance} which highly ranks nodes with high degree that also have neighbors of high degree; these nodes are called ``hubs.'' Once the nodes in both networks are ranked by Importance, the alignment is seeded with the largest hubs and extended greedily by Importance; when the alignment can no longer be greedily extended, a new seed is chosen and the process repeats.  We explicitly coded the Importance measure into SANA so it can be directly optimized. Importance is small compared to other similarity measures; a typical Importance score for an alignment can be smaller than $0.01$. Because of this, it was necessary to weight Importance much more heavily than EC or $S^3$ to balance the objective functions effectively.

{\bf PROPER} generates a group of seed pairs using high-scoring BLAST \citep{blast} bitscores and simultaneously extends these seeds in a greedy fashion. PROPER is one of only two algorithms (the other being NATALIE 2.0) that {\em absolutely require} sequence similarities to run. We would argue that by requiring sequence, PROPER is shooting itself in the foot because it is known that function (the goal of network alignment) correlates with topology \citep{topo-function,graal}, and there are cases of high functional similarity in the absence of sequence similarity \citep{functionsequence}. Aligners that insist on sequence similarity will completely miss the opportunity to align proteins with functional but not sequence similarity. Quoting their paper, PROPER ``chooses the next couple in a greedy way: it chooses the couple with the maximum number of common neighbors \ldots and permanently aligns them'' \citep[Section ``PROPER: two steps'']{proper}.  We note that, given a complete alignment, this score is {\em exactly} twice the numerator of EC (i.e. the number of edges that ``overlap'' between the two graphs in an alignment), since the ``local'' EC centered at an aligned pair is the number of common neighbors, and each such edge is counted twice.  Thus, PROPER's ``graph topology'' is simply a greedily-built EC, and so there is no need to introduce a new topological objective; to compare against them we simply have SANA optimize EC.

{\bf GHOST} relies on a two step process to optimize its new objective, spectral signature similarity. First, it uses a classic seed and extend algorithm to find a high scoring partial alignment, then switches to a local search procedure to find a good total alignment. Their objective function looks in the immediate area of each node (up to a certain length $k$ away), computes a normalized Laplacian matrix of the induced subgraph, performs eigendecomposition on the matrix, and scores the similarity between nodes in different graphs based on the distribution of the eigenvalues of the nodes. We modified the source code of GHOST to output its similarity matrix so that we could optimize it with SANA for comparison. We note that, on large networks, GHOST takes {\em extreme} amounts of memory and CPU to compute its spectral measures---the HSapiens from BioGRID \citep{biogrid_networks} took 11 days utilizing 64 cores and over 120 GB of RAM. However, spectral signatures need only be computed once per network and stored; after that, GHOST quickly computes the similarities between networks and then produces an alignment efficiently.

{\bf WAVE} (Weighted Alignment VotEr) made three important innovations: first optimizing both a node cost and and edge cost function simultaneously, second doing a level-playing-field comparison of two objective functions, and third introducing the idea of {\em weighted} EC or ``WEC.'' It uses a seed-and-extend approach to build an alignment using both node and edge conservation measures.  We have programmed the WEC used by WAVE directly into SANA in order to compare SANA to WAVE. We compared against WAVE using GDV-WEC as recommended by their paper.

\subsubsection{Lagrangian Relaxation} \hfill

{\bf NATALIE 2.0} looks at alignment as a special case of the quadratic assignment problem, and builds its alignments in an attempt to maximize topology using Lagrangian relaxation. However, the search space is exponentially too large, so NATALIE 2.0 {\em requires} us to restrict the alignment possibilities by first filtering which pairs are ``likely'' based upon sequence. Like PROPER, NATALIE 2.0 ``cheats'' in that it requires sequence similarities to prune the search space, even though this defeats the purpose of discovering topology-based functional similarity.

NATALIE 2.0's node-pair objective function is initially based upon sequence similarity and then is updated on each iteration by a Lagrangian relaxation; it can also optimize topology based on EC, $S^3$, or ICS. We told NATALIE 2.0 to output its final (best) node-similarity matrix, which SANA then optimized (along with $S^3$ and EC) for comparison.

{\bf L-GRAAL} is the latest from the GRAAL family of algorithms. It introduces a new graphlet similarity score as an improvement on previous graphlet degree vector similarity, and like NATALIE it uses integer programming and Lagrangian relaxation to search the alignment space. Unlike NATALIE it uses a smarter method of pruning the search space and does not require sequence, although like all the algorithms here it {\em can} use sequence. To compare against L-GRAAL, we directly programmed L-GRAAL's new graphlet-based objective function into SANA so it can be directly optimized.

\subsubsection{Miscellaneous Search Algorithms} \hfill


{\bf ModuleAlign} uses a unique clustering method to score sequence similarity, but their topological objective function is Importance, taken from HubAlign \citep{hubalign}. They search for good alignments using an iterative application of the Hungarian algorithm \citep{hungarian}, slowly expanding the alignment from an initial, partial alignment created by the Hungarian algorithm. We compared against them in a similar fashion as HubAlign, using a judicious weigthing of EC, S3, and Importance.

{\bf GREAT} (GRaphlet Edge-based network AlignmenT) extends the well-known graphlet degree vector objective function \citep{graal} from node-centered orbits to edge-centered orbits. In particular, GREAT computes an {\it edge graphlet degree vector} (edge-GDV) for every edge in each graph.  Then, analogous to the graphlet-orbit similarity, GREAT computes a {\it edge-GDV-similarity} between every pair of edges $(e_i,f_j)$ for one edge in each network. They also compute an {\em edge centrality} measure called {\it edge-GDC}; together edge-GDV-similarity and edge-GDC are used to define a score for aligning any two edges together. Interestingly, this can be viewed as another form of WEC, in the sense that EC gives a score of exactly 1 to each aligned edge, while GREAT provides a score in [0,1] for each aligned edge. Unlike WAVE, this WEC has nothing to do with simply averaging the scores of the endpoint nodes-pairs. They then ``align'' edges, initially without regard to connectivity, using either a greedy (fast) or Hungarian (slow) algorithm.  However, this edge alignment may have low connectivity, so their final stage creates a node-based cost function and produces their final alignment using the node similarities.  As with the edge stage, the node-based alignment can be created using a greedy or Hungarian approach, the latter being far more expensive and producing only marginally better alignments according to their tests.

We think it would be more interesting to create an alignment based directly upon the weighted EC value produced by their edge-GDV-similarity. Thus, we used GREAT to produce the (very large) edge-GDV-similarity matrix that provides a score to every pair of edges (one from each graph), and then we used SANA to directly optimize the resulting weighted EC. Unfortunately, GREAT takes far to long to create the edge-GDV's for networks with large amounts of edges, about three months wall-clock time, so we did not run GREAT on the SCerevisiae or HSapiens networks. Furthermore, running the second half of GREAT using the Hungarian algorithm also takes around two to three months wall-clock time, so we ran GREAT with the first half in Hungarian mode and the second half in greedy mode (what they call ``G-H''). We compared against GREAT by running their algorithm with an 80-20 split between edge-GDV-similarity and edge-GDC-similarity because their paper reported best results for this weighting.

\subsection{General Testing Details and Datasets}

As mentioned in the paragraph about SANA, we added $S^3$ and EC to SANA's objective function to ensure connectivity. For graphs in section \ref{subsec:maingraphs} (The first two columns of Figure \ref{fig:charts}), unless otherwise stated, we show how SANA does with a ``balanced'' objective function where multiple objectives are weighted (e.g. 30\% EC, 20\% $S^3$, and 50\% GDV Similarity). This allows SANA to perform better than other algorithms at EC, $S^3$, and the algorithm's objective function (from Table \ref{tab:aligners}) simultaneously. All objective function weightings can be seen in Table \ref{tab:balancing}.

We later show in section \ref{subsec:splitgraphs} compiled graphs of how well SANA performs at other alignment algorithms' objective functions with balanced objective functions and depict in section \ref{subsec:allgraphs} graphs of how SANA performs when \textit{only} optimizing the objective function of the other aligner---without balancing the objective function with EC nor $S^3$. It should be noted that all scores for a particular pair of networks for a particular objective function represent the scores of only {\em one alignment}, not a combination of the best scores from SANA and the other aligner.

We compared SANA to the other algorithms on the 8 BioGRID networks, which is a total of 28 pairs between \textit{Rattus norvegicus}, \textit{Schizosaccharomyces pombe}, \textit{Caenorhabditis elegans}, \textit{Mus musculus}, \textit{Saccharomyces cerevisiae}, \textit{Arabidopsis thaliana}, \textit{Drosophila melanogaster}, and \textit{Homo sapiens}. For several aligners, we were forced to compare against a smaller set: GREAT and NATALIE 2.0 failed to run for any pair involving HSapiens or SCerevisiae, and WAVE failed to run for any pair involving HSapiens.

\section{Results}

\subsection{Individual Comparisons}
\label{subsec:maingraphs}

For each pair of networks from BioGRID \citep{biogrid_networks}, we ran all of the network aligners described in section \ref{subsec:comparedalgs} against SANA, where SANA optimized EC, $S^3$, and the objective function of the other aligner. Since SANA almost universally outperforms the other algorithms in all measures, we introduce the idea of the \textit{SANA Improvement Factor}, or \textit{SIF}, which is the factor by which SANA outperforms the other aligner on any given measure:\[SIF=\dfrac{SANA's\ Score}{Other\ Algorithm's\ Score}.\]

Since almost all of the objectives lie in [0,1] and the SIF is almost always greater than 1, a convenient way to plot all the scores on one graph is to plot the objective values for $y<1$ and the SIF for $y>1$. This gives 6 curves for each plot: the raw scores the other aligner got for EC, $S^3$, and objective function of the other aligner below $y=1$, and the three SIFs above $y=1$. It should be noted that the top half of the graphs (where the SIFs generally lie) does not necessarily have the same scale factor as the bottom half of the graphs, because the SIF is often much larger than 1. These charts follow the key in Figure 1.

\begin{figure}[h]
\centering
\includegraphics[width=4cm]{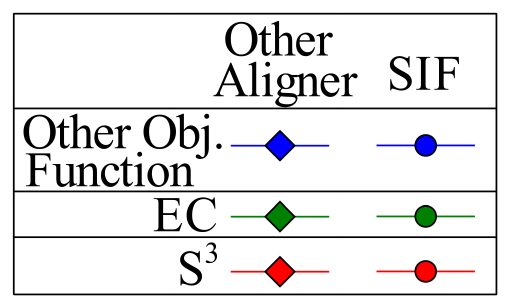}
\caption{Key for the first two columns of Figure \ref{fig:charts}.}
\label{fig:key1}
\end{figure}

\begin{table*}
\begin{center}
\caption{SANA Objective Weights by Network Aligner}
\label{tab:balancing}
\begin{tabular}{|| c c c ||} 
 \hline
 Alignment Algorithm & Objective Function & SANA Weights  \\ 
 \hline \hline
 MAGNA++ & GDV Similarity & 10\% $S^3$, 90\% GDV Similarity \\ 
 \hline
 OptNetAlign & $S^3$ & 100\% $S^3$\\ 
 \hline
 HubAlign & Importance & 1.992\% EC, .008\% $S^3$, 98\% Importance\\
 \hline
 GREAT & Edge GDV Similarity & 50\% $S^3$, 50\% Edge GDV Similarity\\
 \hline
 PROPER & (Effectively) Edge Coverage & 50\% EC, 50 \% $S^3$\\
 \hline
 GHOST & Spectral Signature Similarity & 30\% EC, 30\% $S^3$, 60\% Spectral Signature Similarity\\
 \hline
 WAVE & WEC Weighted by GDV Similarity & 20\% $S^3$, 80\% GDV-Weighted WEC \\
 \hline
 NATALIE 2.0 & Natalie's Objective Function (Unnamed) & 25\% EC, 25\% $S^3$, 50\% Natalie's Obj. Func.\\ 
 \hline
 L-GRAAL & L-GRAAL GDV Similarity & 5\% $S^3$, 95\% L-GRAAL GDV Similarity \\
 \hline
 ModuleAlign & Importance & 1.992\% EC, .008\% $S^3$, 98\% Importance\\
 \hline
\end{tabular}
\end{center}
\end{table*}

\begin{figure*}[htbp]
\centering
\includegraphics[width=.33\textwidth]{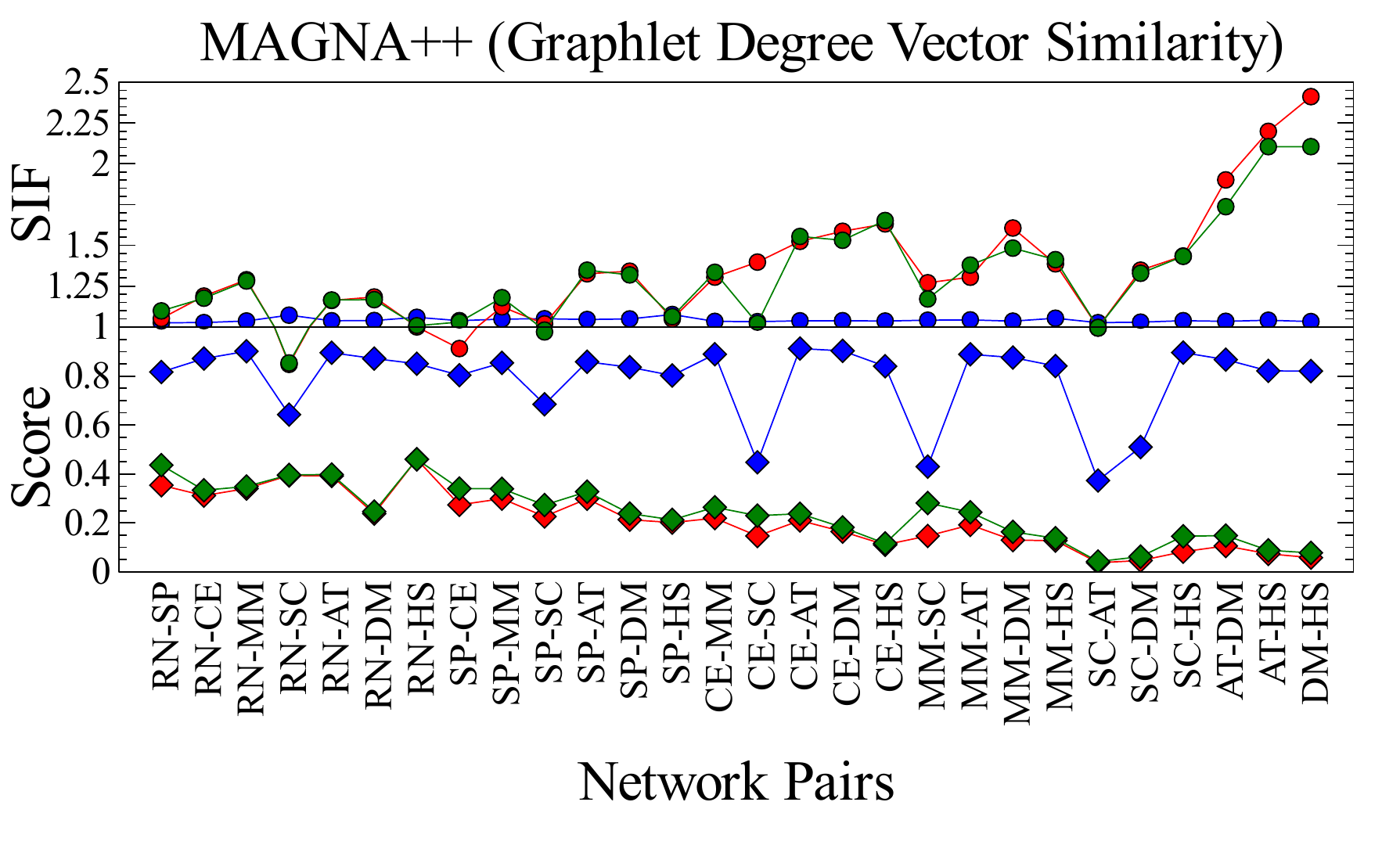}
\includegraphics[width=.33\textwidth]{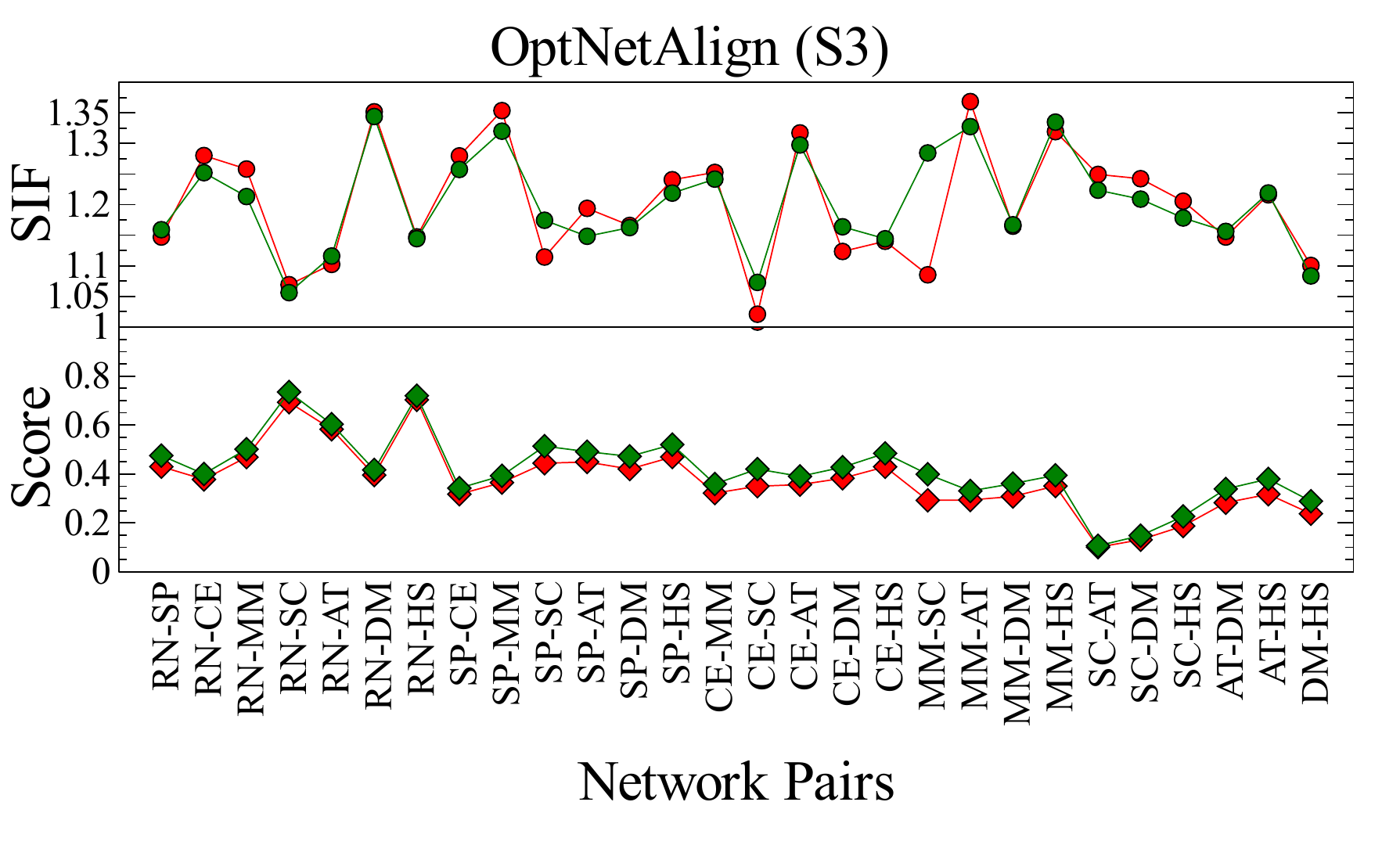}
\includegraphics[width=.33\textwidth]{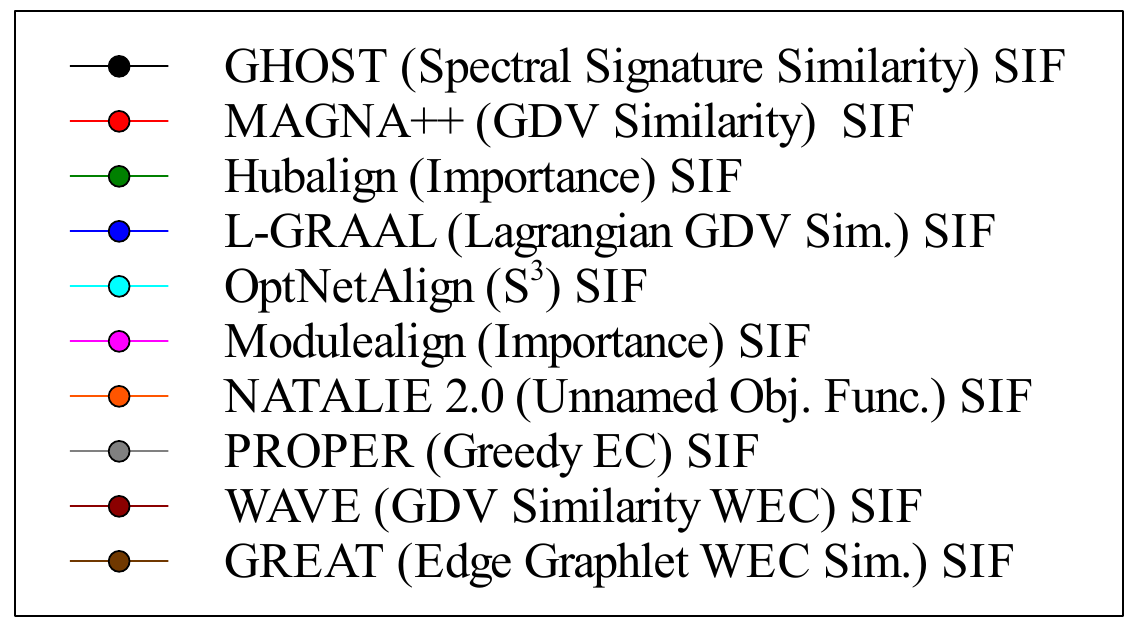}\\
\includegraphics[width=.33\textwidth]{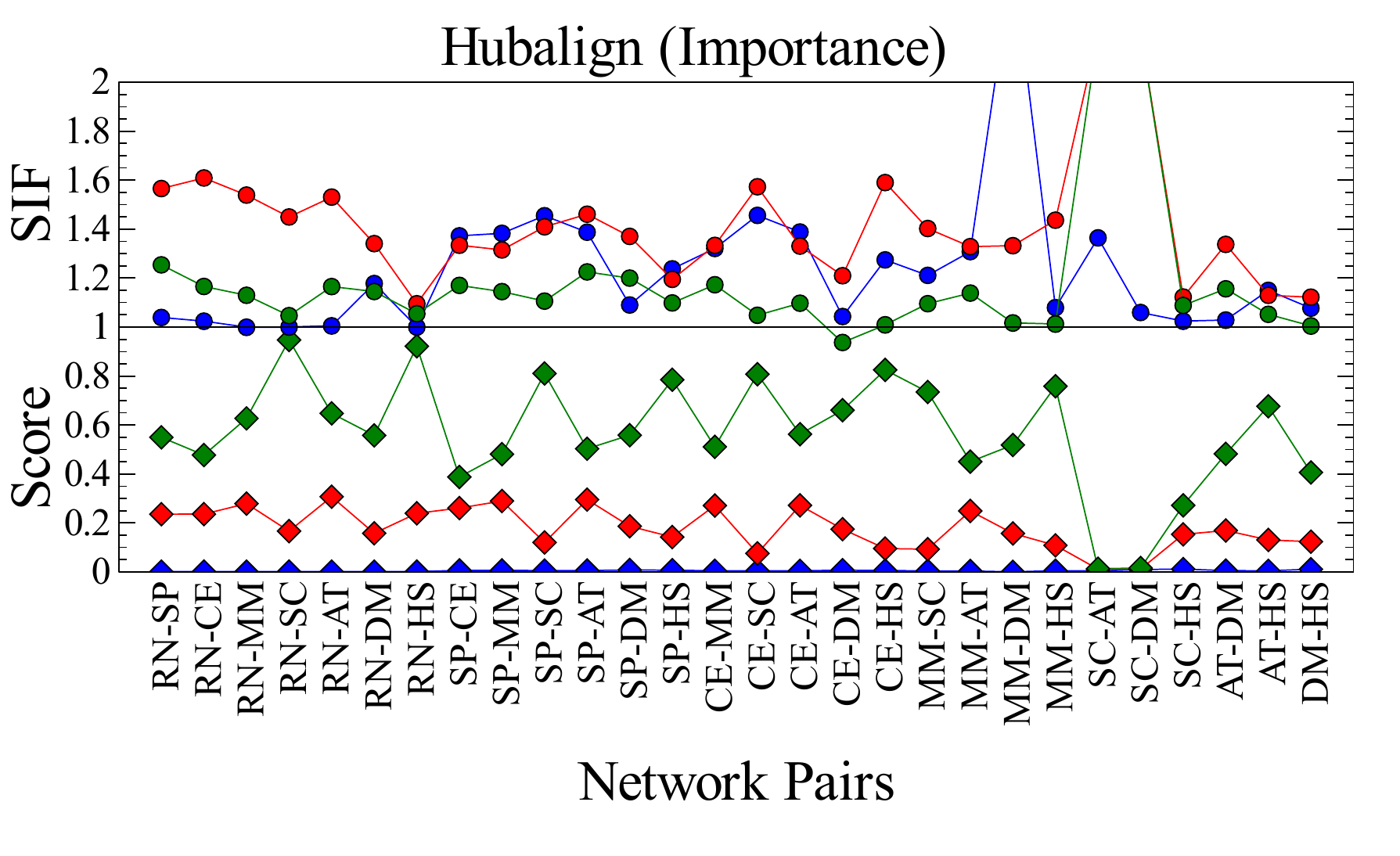}
\includegraphics[width=.33\textwidth]{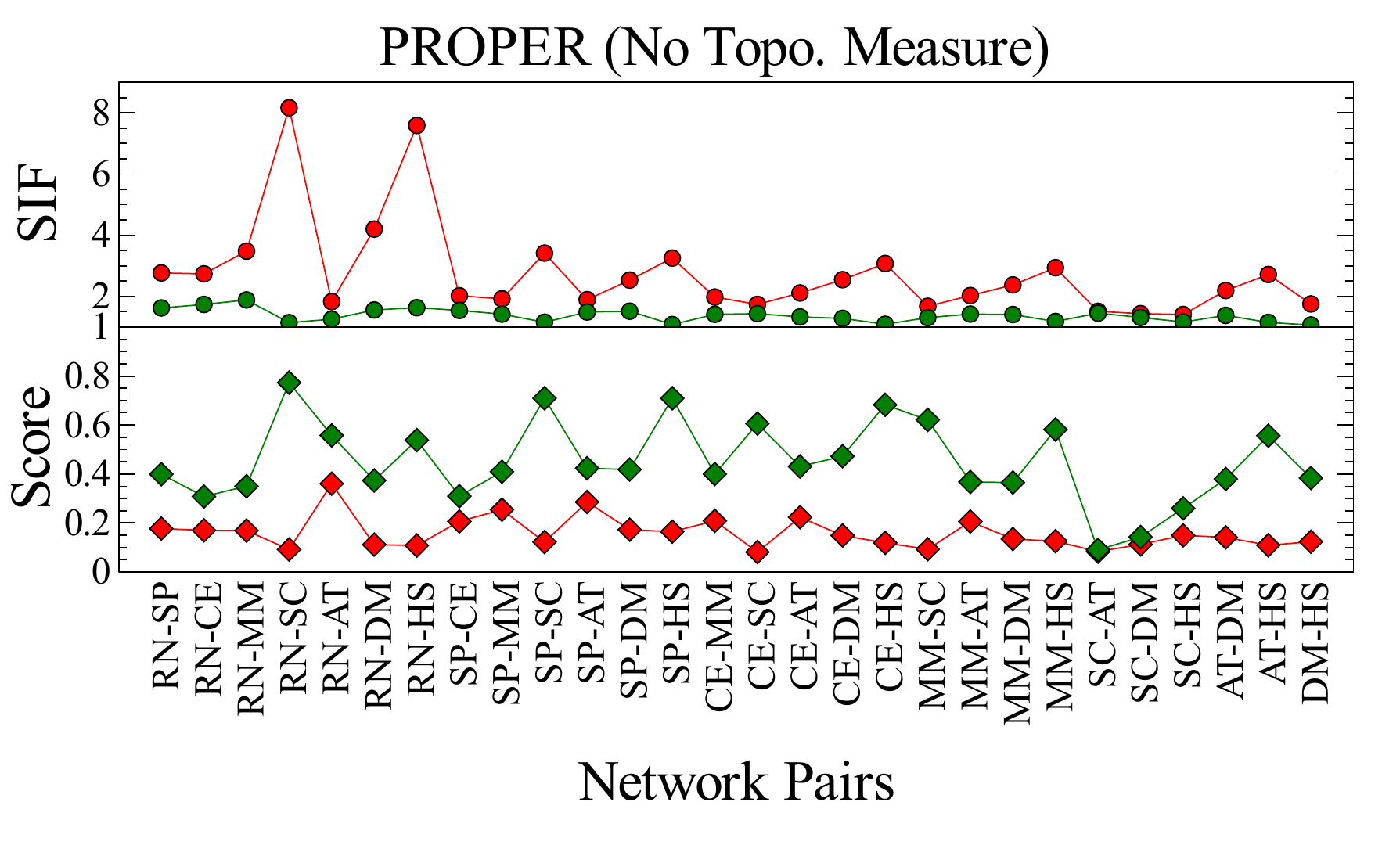}
\includegraphics[width=.33\textwidth]{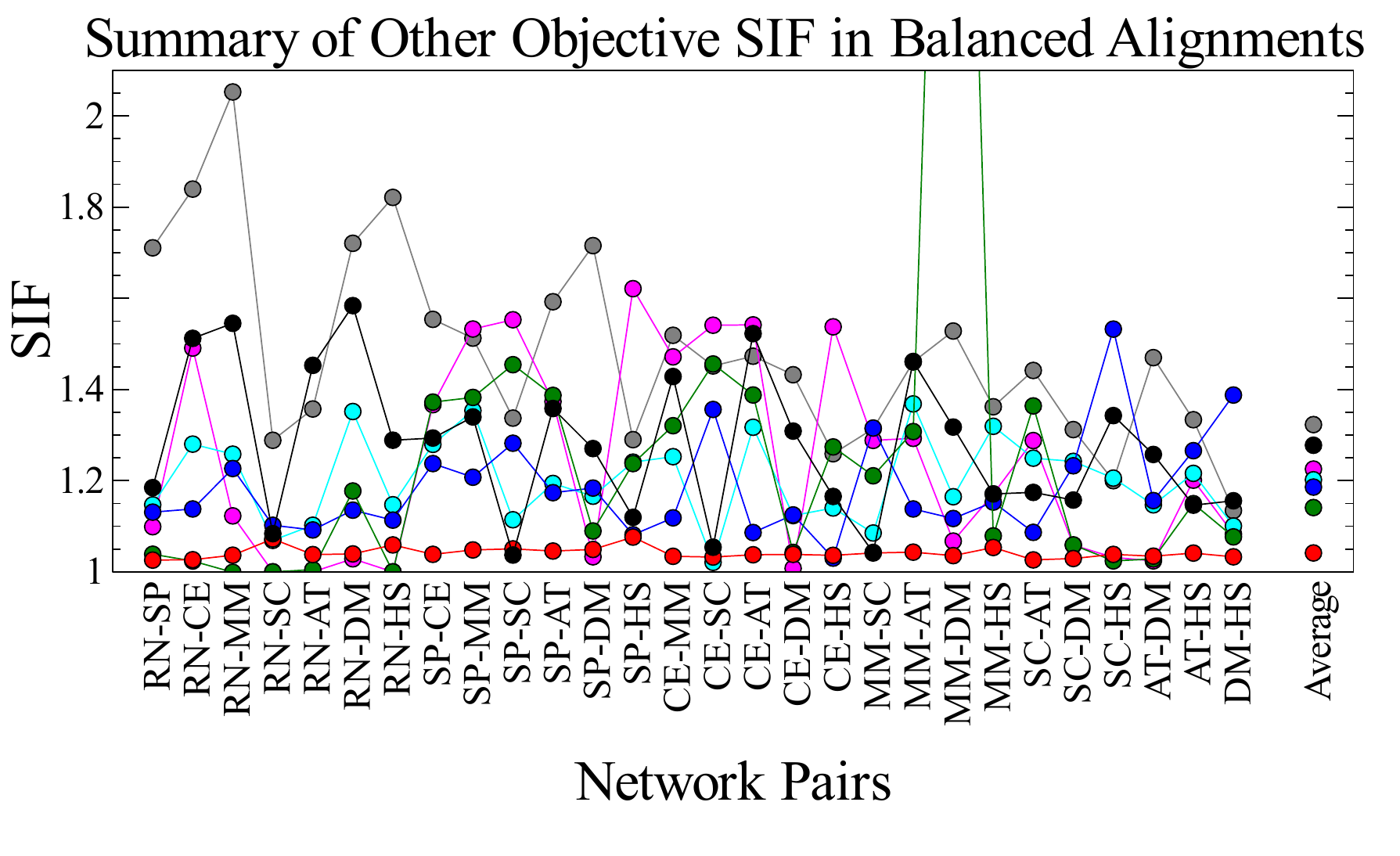}\\
\includegraphics[width=.33\textwidth]{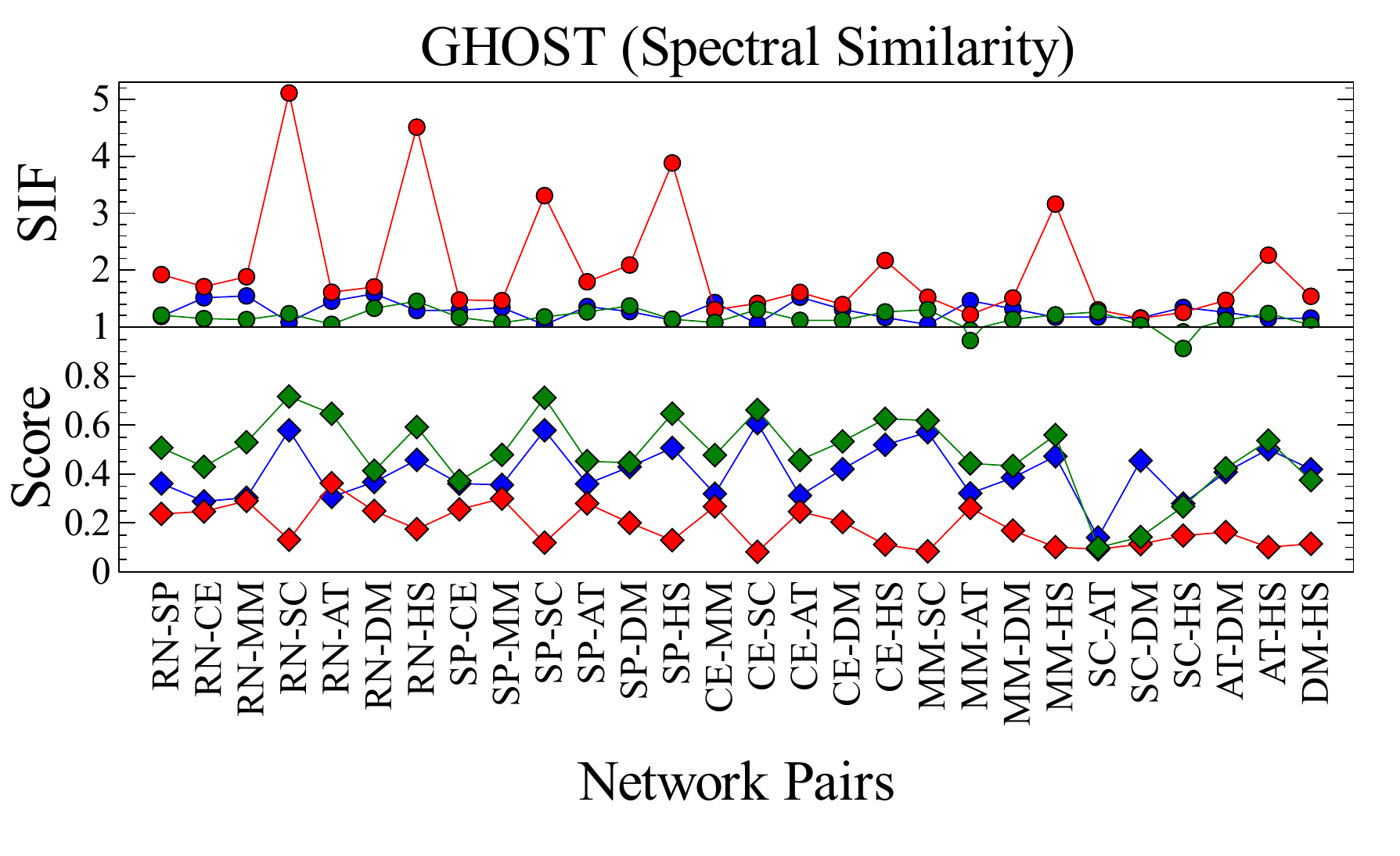}
\includegraphics[width=.33\textwidth]{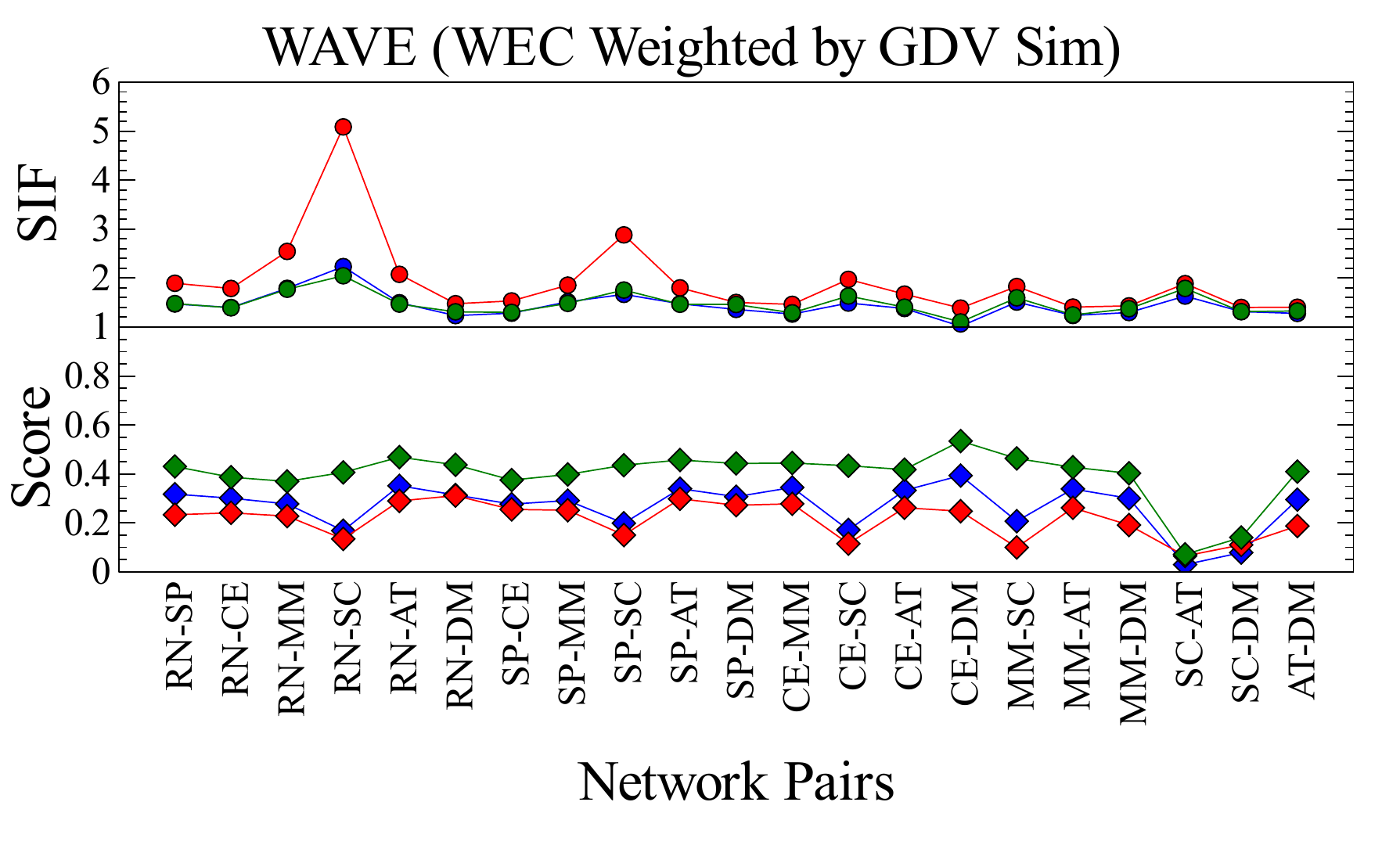}
\includegraphics[width=.33\textwidth]{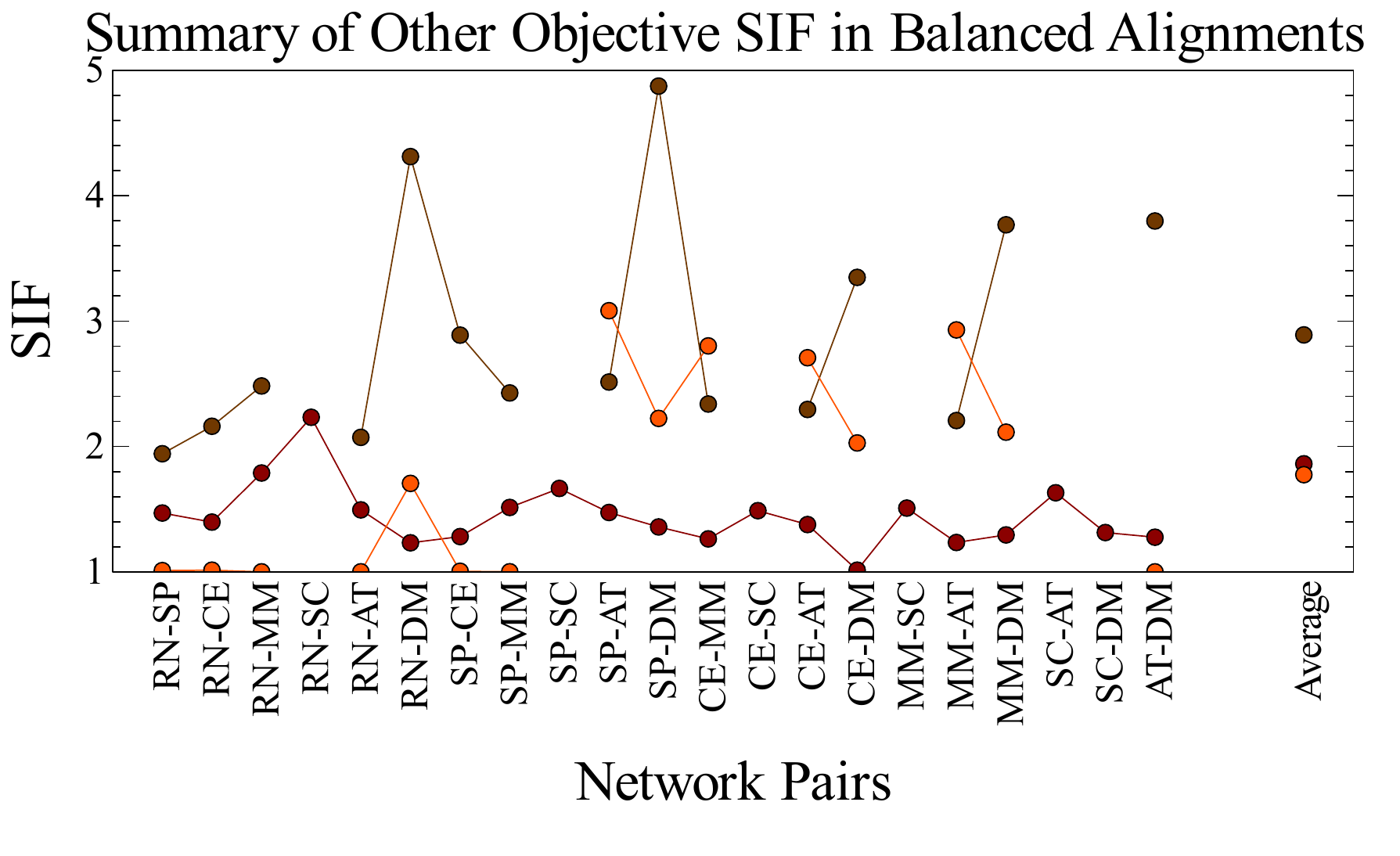}\\
\includegraphics[width=.33\textwidth]{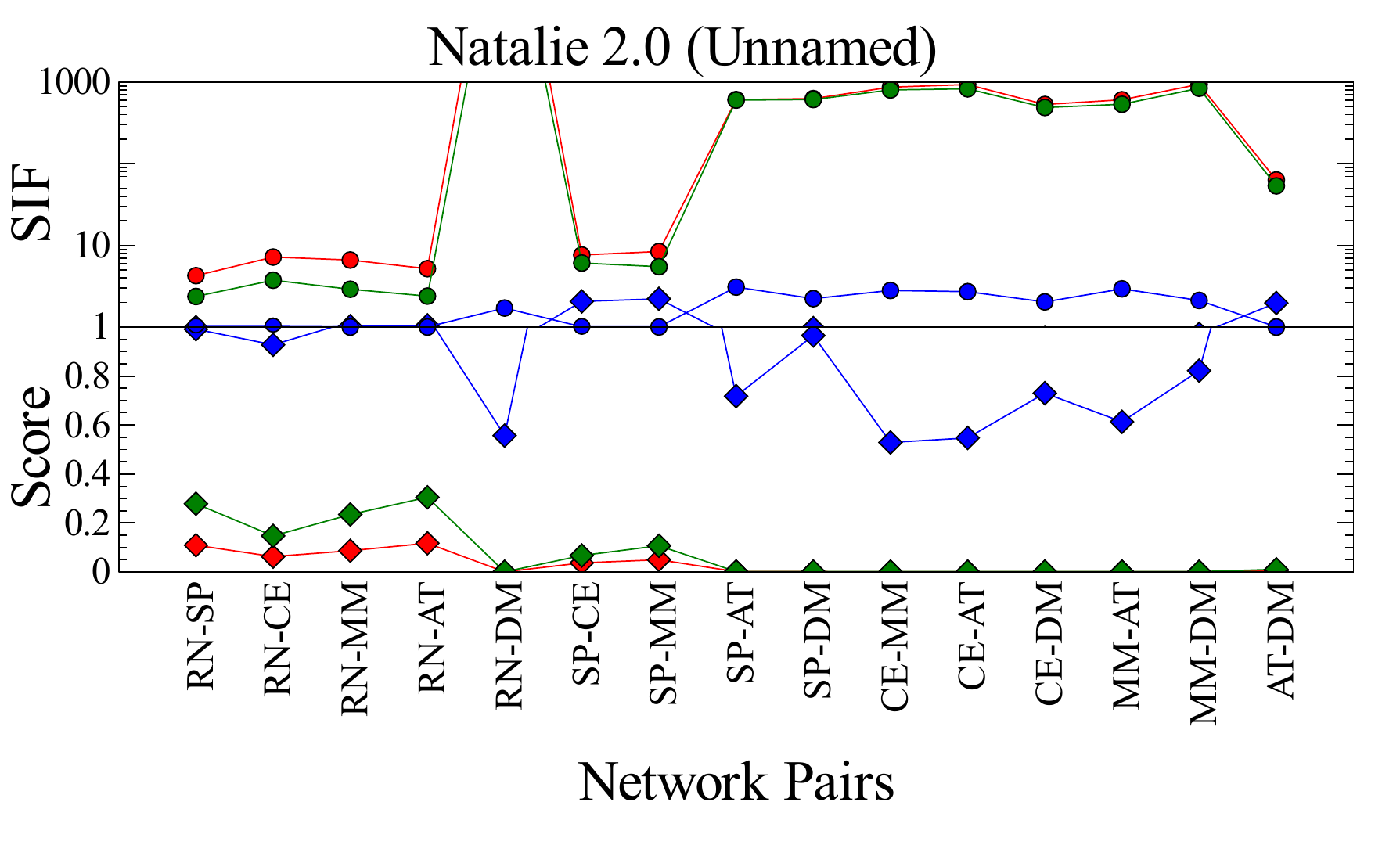}
\includegraphics[width=.33\textwidth]{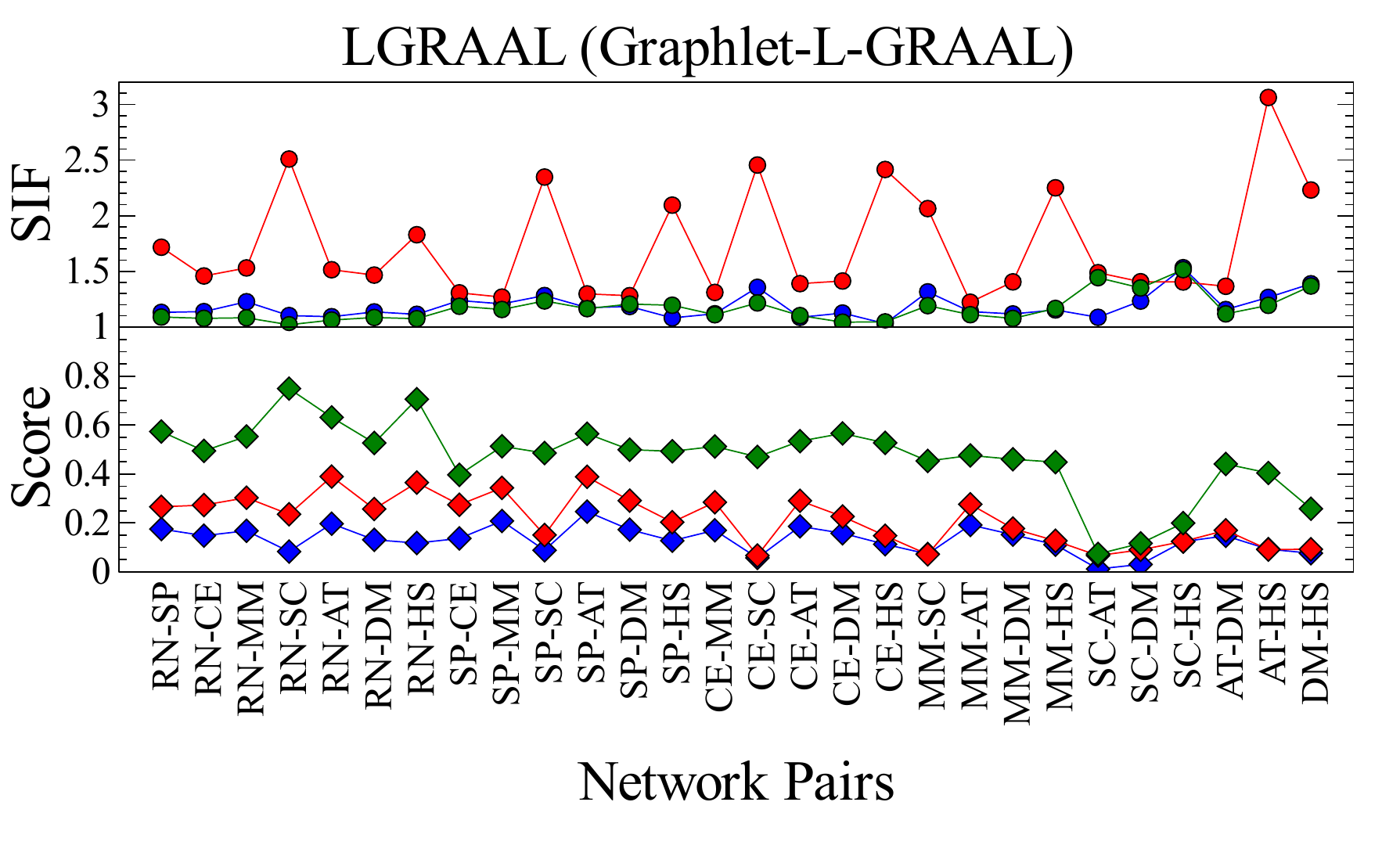}
\includegraphics[width=.33\textwidth]{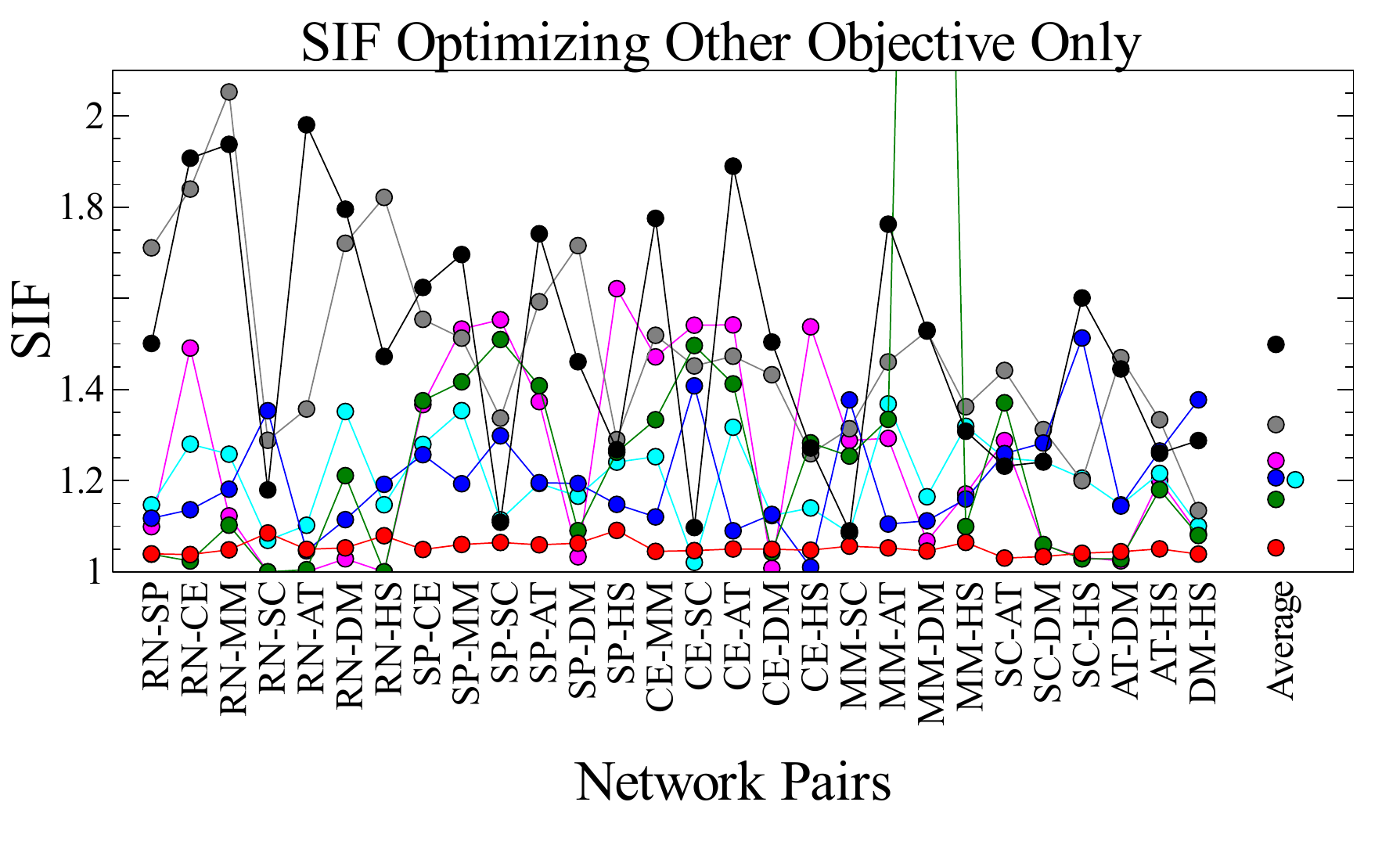}\\
\includegraphics[width=.33\textwidth]{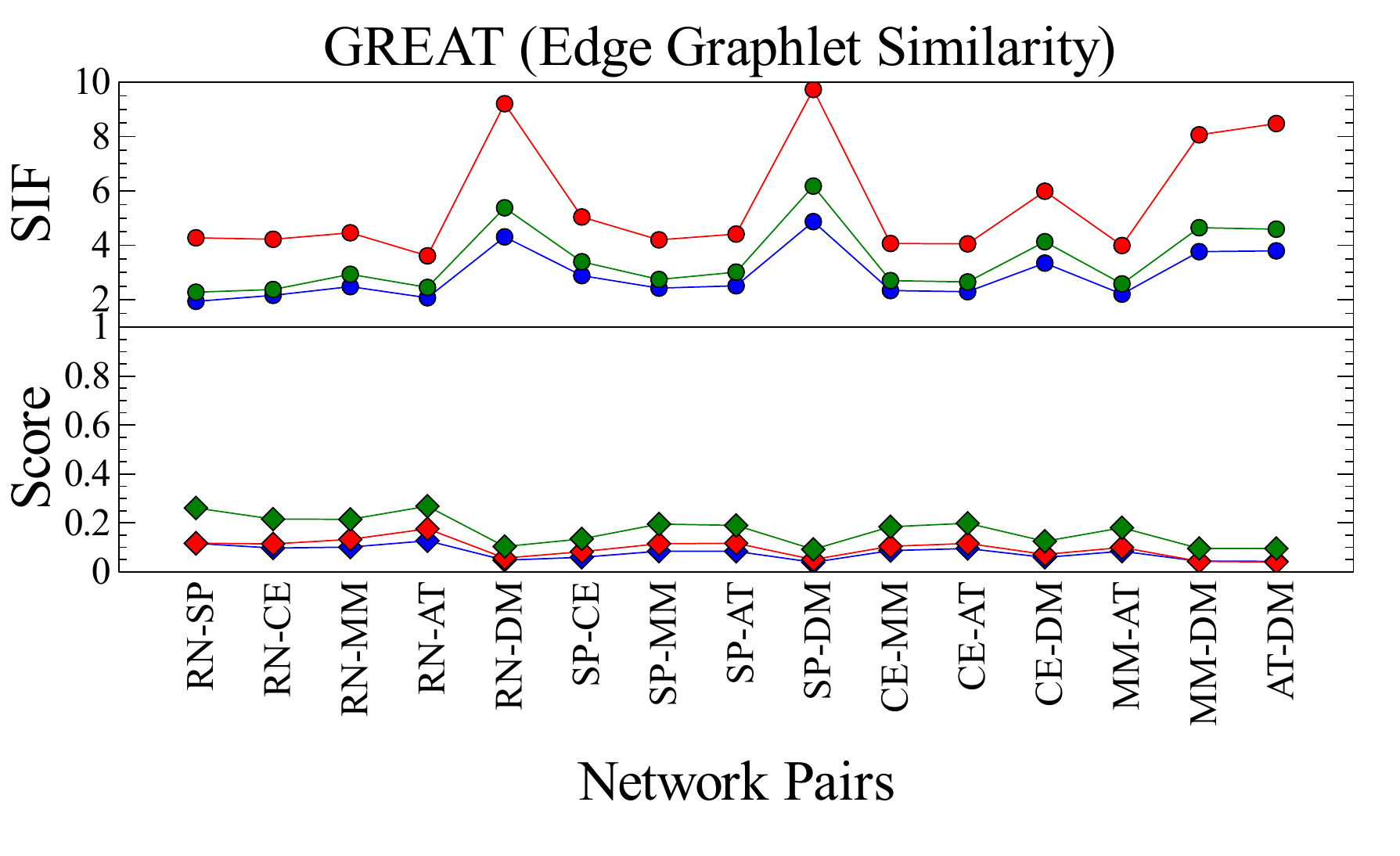}
\includegraphics[width=.33\textwidth]{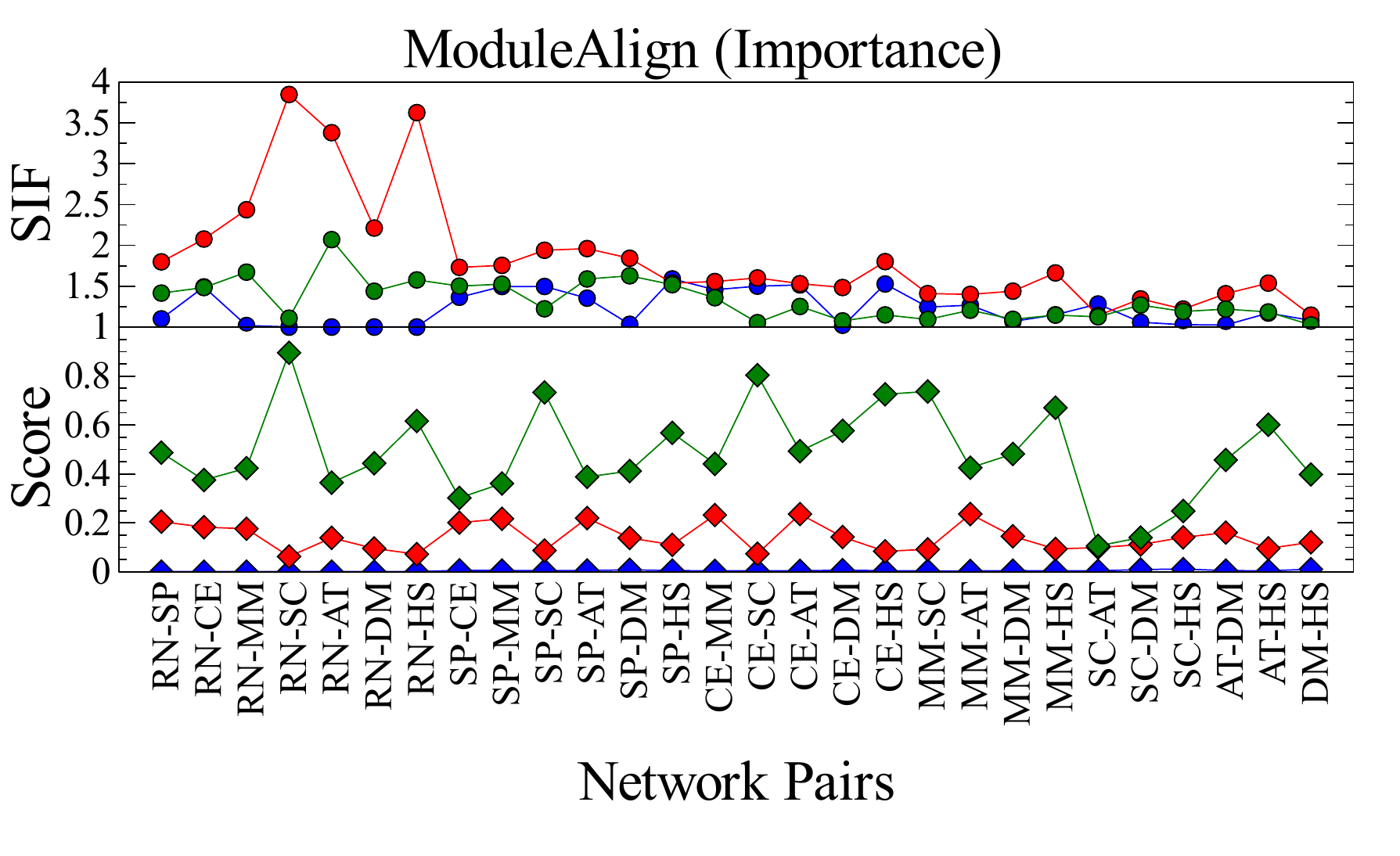}
\includegraphics[width=.33\textwidth]{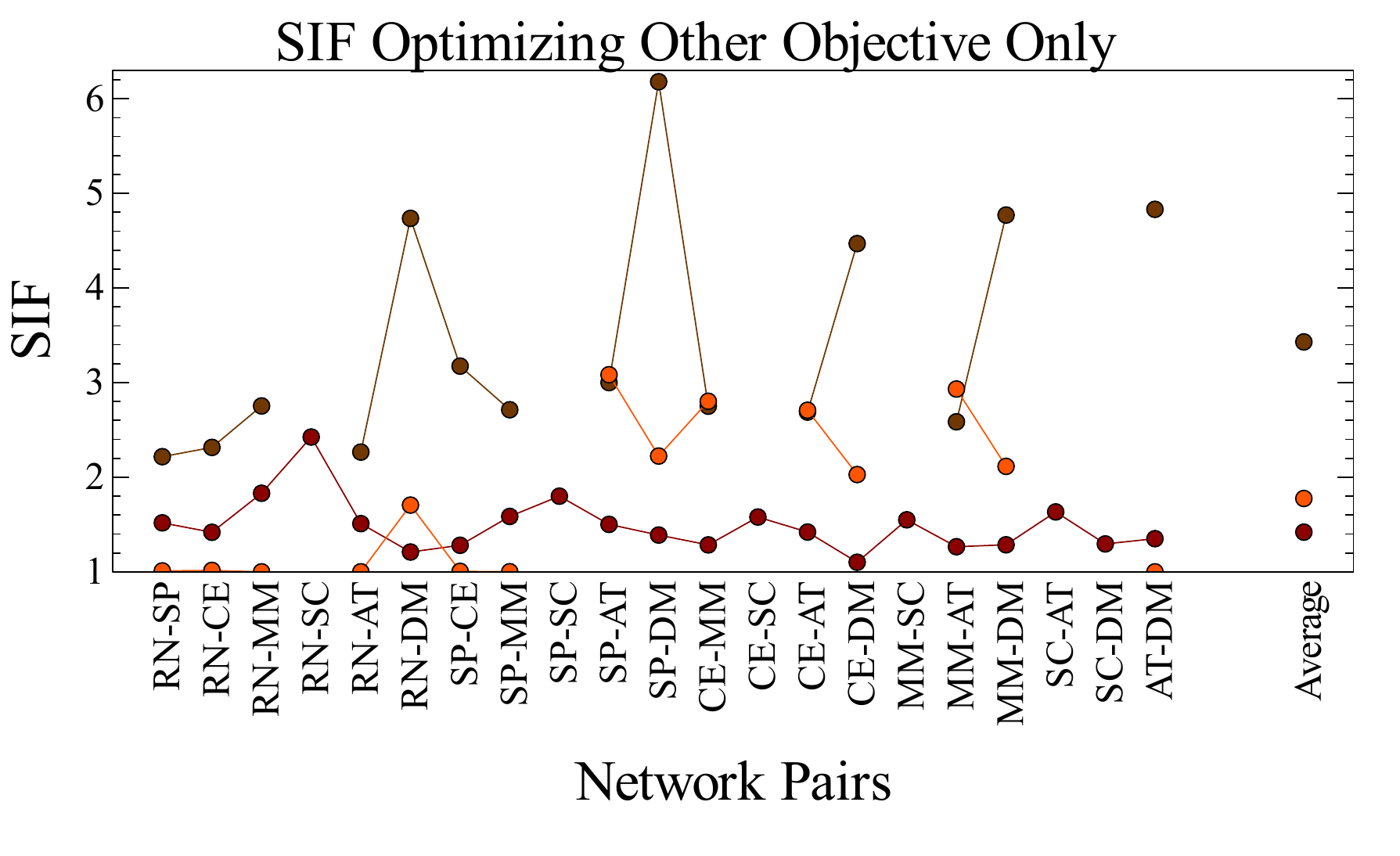}
\caption{All comparison charts for 20-minute runs of SANA. The first two columns follow the key in Figure \ref{fig:key1}, while the last column uses the key at the top of the column. For each plot, the horizontal axis enumerates pairs of BioGRID networks. In each plot, the curves below the $y=1$ line depict the competing aligner's score for EC (green diamond), S3 (red diamond), and its own objective function (blue diamond). The curves above 1 depict the ``SANA Improvement Factor'' (SIF) by which SANA outscores the competing aligner for the same measures. Note that all of SANA's curves are virtually always above or equal to 1, depicting how SANA almost always outperforms the competing aligners. The ``balanced'' weights used by SANA when optimizing the combined objectives are depicted in Table \ref{tab:balancing}. The third column follows the key at the top of the column and depicts only the SIF for all aligners when either optimizing ``balanced'' alignments or alignments where SANA only optimized the other objective function.}
\label{fig:charts}
\end{figure*}

The first two columns of Figure \ref{fig:charts} depict individual comparisons of SANA to each of the other aligners. A few important details should be explained about these graphs in particular. First, any scores that go above the top of a graph were too large to easily present on the graph. Some of these scores approach infinity, because the other network alignment algorithm got a score of zero for that particular measure, such as NATALIE 2.0's EC and $S^3$ score when aligning RNorvegicus to DMelanogaster. Some of the scores above the graph were finite, but their exact value is not as important as the underlying point: SANA vastly outperformed them in that circumstance.

Also, note that half of the y axis for the NATALIE 2.0 comparison graph is logarithmic. We made this graph logarithmic because it most easily fit all of the data points. There is some very small visual duplication of certain data points in the NATALIE 2.0 graph because the scales for the top and bottom half of the graphs are wildly different, and NATALIE 2.0's objective function can be higher than 1.

All of the results depicted in Figure \ref{fig:charts} are for 20 minute runs of SANA. As a more stringent test, we also performed the exact same tests for even shorter runs of SANA of 3 and 8 minutes.
We find that across all aligners and all pairs of networks, and assigning the weights depicted in Table \ref{tab:balancing}, SANA beats the other aligners 82.8\% of the time for 3 minute runs of SANA, 94.5\% for 8 minute runs, and 97.3\% for 20 minute runs; these figures are a lower bound on SANA's superiority because we used the same weighting across all network pairs, for each aligner; presumably we could do even better if we were to optimize the weights for every network pair individually.

\subsection{Summary of Individual Comparisons}
\label{subsec:splitgraphs}

The first two charts in the last column of Figure \ref{fig:charts} indicate a compilation of the SIF the other network aligners' objective functions (all of the blue circle curves) from the charts in the first two columns. Clearly, because the SIF for all other objective functions on the charts is greater than or equal to 1, SANA is more than capable of optimizing other aligners' objective functions better than they can. The top of the last column of Figure \ref{fig:charts} also includes a new key for the rest of the column. There are two graphs to summarize the first two columns because, as mentioned earlier, some aligners did not run for all pairs of BioGRID networks.

\subsection{SANA Optimizing Only Outside Objective Functions}
\label{subsec:allgraphs}

We also compared SANA when optimizing only the objective function of the other alignment algorithms without adding EC and $S^3$ weights. We include these SIF scores in the last two charts of the third column of Figure \ref{fig:charts}, and the key for the figure can be seen at the top of the same column. Again, none of the SIF values are below 1. These graphs show how well SANA can perform at its best without concern for compensating with topological connectivity. These comparisons are not as fair as the previous comparisons because SANA did not enforce topological connectivity on these alignments with EC and $S^3$.

\section{Discussion}

\subsection{Complete Comparison Summary}

As seen in section \ref{subsec:maingraphs} and \ref{subsec:splitgraphs} (first two columns and first two charts in third column of Figure \ref{fig:charts}), SANA soundly outperforms the other algorithms at their own objective functions. Section \ref{subsec:allgraphs} further demonstrates that when SANA optimizes \textit{only} the other objective function without adding other topological measures, it often vastly outperforms the other alignment algorithms at their own objective function.

Some aligners do well against SANA, performing only slightly worse than SANA in most cases, like MAGNA++ and OptNetAlign (Figure \ref{fig:charts}). We suspect this is because these two algorithms rely on random searches instead of deterministic methods that allow it to view more of the search space of possible alignments rather than being restricted by quality criteria. As discussed in the original SANA paper \citep[section 2.1]{sana1}, exploring worse alignments allows these algorithms to escape local score maxima. MAGNA++ fares especially well in this regard only getting beaten by about 5\% by SANA in its own measure, GDV-Similarity.

Some of the worst performing algorithms were seed-and-extend algorithms, like GREAT, PROPER, GHOST, and WAVE. SANA significantly outperformed them at their own objective function, especially GREAT. The only exception to this pattern was HubAlign, which often scored equally well ($SIF=1$) compared to SANA, suggesting ``Importance'' is an aptly named measure. Nevertheless, we believe the superiority of SANA, MAGNA++, and OptNetAlign solidifies our argument that random search is superior at more thoroughly exploring the search space of possible alignments.

\subsection{Triangle Alignment}

We relegate TAME \citep{tame}, another network aligner that we compared against, to the supplementary because TAME considers {\em only} triangles in its alignments; this is problematic for reasons we discuss in the Supplementary; for now we simply observe that some BioGRID networks have as little as 20\% of their edges in triangles, and we find it difficult to believe that any algorithm that completely ignores 80\% of the edges in a network could possibly provide good alignments. Furthermore, it has been shown that more than a dozen other graphlets are necessary to capture topology-function relationships in PPI networks \citep{topo-function}.

\subsection{Simulated Annealing's Effectiveness}
\label{subsec:temperature}

Simulated annealing has proven to be quite effective at optimizing explicit objective functions. As mentioned earlier, one of the keys to SANA's speed and effectiveness is ``incrementally evaluating'' the objective functions so they don't need to be entirely recomputed each time SANA changes the alignment. This makes computing objective functions that normally require checking every edge or every node in both networks, like Edge Coverage, orders of magnitude faster, because only a very small subset of the total information is necessary to recompute the score. Incrementally evaluating objective functions is critical to quickly anneal. SANA can perform over a million iterations per second on most machines, but without incremental evaluation it slows by orders of magnitude. OptNetAlign also speeds up their alignment algorithm using incremental evaluation.

Furthermore, SANA also has a method of automatically generating the temperature schedule so the run progresses smoothly from accepting most random moves to accepting no random moves. This is another critical part of SANA's effectiveness; a bad temperature schedule means the annealing will either make mostly random moves (causing no progress to be made increasing the objective function) or make far too few random moves (causing the anneal to stagnate in local maxima). A good temperature schedule can be determined by hand with some empirical measurements and math, but it is completely impractical to perform multiple times, especially not when hundreds of unique temperature schedules are needed (such as in this paper). By automatically finding a good temperature schedule we speed this process up and make simulated annealing practical. We will describe the details of the automatic temperature schedule in future work.

We feel that incrementally evaluating the objective functions and automatically determining the temperature schedule are the key discoveries we made that bring out the full power of simulated annealing in SANA. This allows SANA to be the most effective at optimizing almost any explicit objective function. There is one other simulated annealing algorithm for network alignment (\cite{sailmcs}). Since it is primarily designed for multiple network alignment, we defer a detailed comparison with it until SANA is capable of aligning multiple networks.  For now, we note that when aligning just 2 networks, SailMCS optimizes EC, and across the BioGRID network it achieves a mean EC of 0.24 compared to SANA's 0.61.


The main ability that SANA does not possess is the ability to implicitly optimize general qualities that are not easily quantified in explicit objective functions. For example, when SANA optimizes \textit{only} Hubalign's objective function, Importance, the resulting alignments have very poor EC, $S^3$, and general connectivity. Nevertheless, Hubalign's search algorithm, because it is greedy and follows edges, creates alignments with good topology. The exact nature of how Hubalign's alignments have good topology is not fully explored or explained. If it turns out, by luck, that another network alignment algorithm implicitly creates good topology because of its search algorithm, SANA cannot reproduce this without some sort of explicit (i.e. mathematical expression) definition of the topology. We do not suspect any current network aligners have "hidden topology" that is necessarily advantageous over current explicit objective functions, but if that is the case SANA may not be the best way to reproduce it. SANA cannot predict what is and is not a good measure; it can only optimize what it is given. We will attempt to, in part, determine the quality of the measures described in this publication in our companion paper \cite{companion}.

\subsection{Biological Significance}

As mentioned in the introduction, there is little biology in this paper because it serves as the foundation for our companion paper \cite{companion} which uses SANA to compare the relative effectiveness of the objective functions with regard to biology. This work serves to prove that SANA is the current best standard to optimize any objective function, so comparisons of objective functions using SANA are fair and without the confounding variable of multiple search algorithms.


\subsection{Conclusions}

The results from section \ref{subsec:maingraphs} and \ref{subsec:splitgraphs} highlight the importance of introducing network alignment algorithms and objective functions separately. Because SANA was not introduced with a unique objective function, implicit topology (like edge-following greedy algorithms), nor the restriction to only optimize one objective function, SANA could be easily compared to other alignment algorithms using any objective function. At the least, alignment algorithms should be introduced without new objective functions so comparisons of their effectiveness can be simple and fair. We explore the idea of comparing the effectiveness of objective functions at reproducing biology in our companion paper \cite{companion} using SANA, because SANA proved to be the most effective network aligner for almost any objective function.

We also believe that the results from section \ref{subsec:maingraphs} and \ref{subsec:splitgraphs} (the first two columns and first two charts of the third column of Figure \ref{fig:charts}) can be extended to indicate that unless a new search algorithm can convincingly beat SANA at reproducing high scoring alignments for multiple objective functions, SANA should be universally adopted as the state-of-the-art network alignment algorithm. SANA can outperform all current network alignment algorithms in a short amount of time for virtually all objective functions.
We encourage work to focus on the creation of new objective functions that can better reproduce topology and biology.

\bibliography{document}{}
\bibliographystyle{natbib}

\end{document}


\firstpage{1}

\title[An Impartial Comparison of Modern Network Alignment Objective Functions Using Simulated Annealing, Part 1: Supplementary Material]{An Impartial Comparison of Modern Network Alignment Objective Functions Using Simulated Annealing, Part 1: Supplementary Material}
\author[Kanne, Hayes]{Dillon Kanne, Wayne B. Hayes\footnote{to whom correspondence should be addressed ({\tt whayes@uci.edu})}\;}
\address{Department of Computer Science, University of California, Irvine CA 92697-3435, USA}

\history{Received on XXXXX; revised on XXXXX; accepted on XXXXX}

\editor{Associate Editor: XXXXXXX}

\maketitle
\begin{abstract}
    In this supplementary material we discuss TAME and its triangle correctness measure as well as give a full chart of the runtimes of all algorithms tested.
\end{abstract}
\section{Triangle Alignment}

Simulated Annealing was built on the assumption of almost every move increasing or decreasing the net energy (score) of a system, just as moving particles in annealing metal increases or decreases the net energy of the metal. In the case of network alignment, this assumes that (almost) every change to the alignment has {\em some} measurable effect on the objective function; there should be some kind of guidance for a large fraction of moves. Some solutions spaces are not like that, but are instead extremely sparse, with many scores of exactly zero interspersed with almost delta-function-like jumps in the score. In such a solution space, almost every move has no effect at all, such as moving from a zero-scoring alignment to another alignment also scoring zero. Even the ``perfect'' objective function, which scores some ``perfect'' alignment with 1 and every other alignment with 0, is impossible to find with random search because each move has no affect on the score; there is no guidance for ``good'' and ``bad'' moves. In such cases, there is little hope of ``converging'' on a good solution since even if one finds oneself temporarily in the vicinity of a delta-function increase in the score, the very nature of the random search (with a non-zero temperature) means we are likely to wander back into zero-score ``flatlands,'' and spend most of our time there. We mentioned these restrictions on simulated annealing in section 2.1.1 of the main paper when introducing SANA.

With the method of moves we have chosen for SANA, one such flatland is ``triangle alignment,'' where the {\em only} thing the objective function cares about is the number of aligned triangles. This is the objective function used in TAME \citep{tame}. With the system of moves we currently use in SANA, we were unable to get a good triangle alignment, for the above reasons. In this sense there can indeed be objectives that SANA may not excel at compared to a hand-coded deterministic algorithm.  We hypothesize that SANA {\em could} easily be modified to do well at aligning triangles.  First, we would exhaustively list all the triangles in each graph, for which asymptotically optimal algorithms exist \citep{Goodrich}; then, we would program SANA's ``moves'' to swap or move entire triangles.  However, we believe this would be a waste of time because the number of triangles in all the biological networks we've encountered involve only a small fraction of all the edges in the networks; we find it hard to believe that {\em any} good alignment algorithm could possibly recover relevant biology by completely ignoring most of the edges in the network.  In the BioGRID networks tested in this paper, the amount of edges participating in triangles ranged from as low as 18\% in the case of CElegans to as high as 92\% in the case of SCerevisiae. Most of the networks have less than half of their edges in triangles, indicating that a {\em majority} of topological information is discarded when using Triangle Conservation. 

\begin{figure}
    \centering
    \includegraphics[width=.5\textwidth]{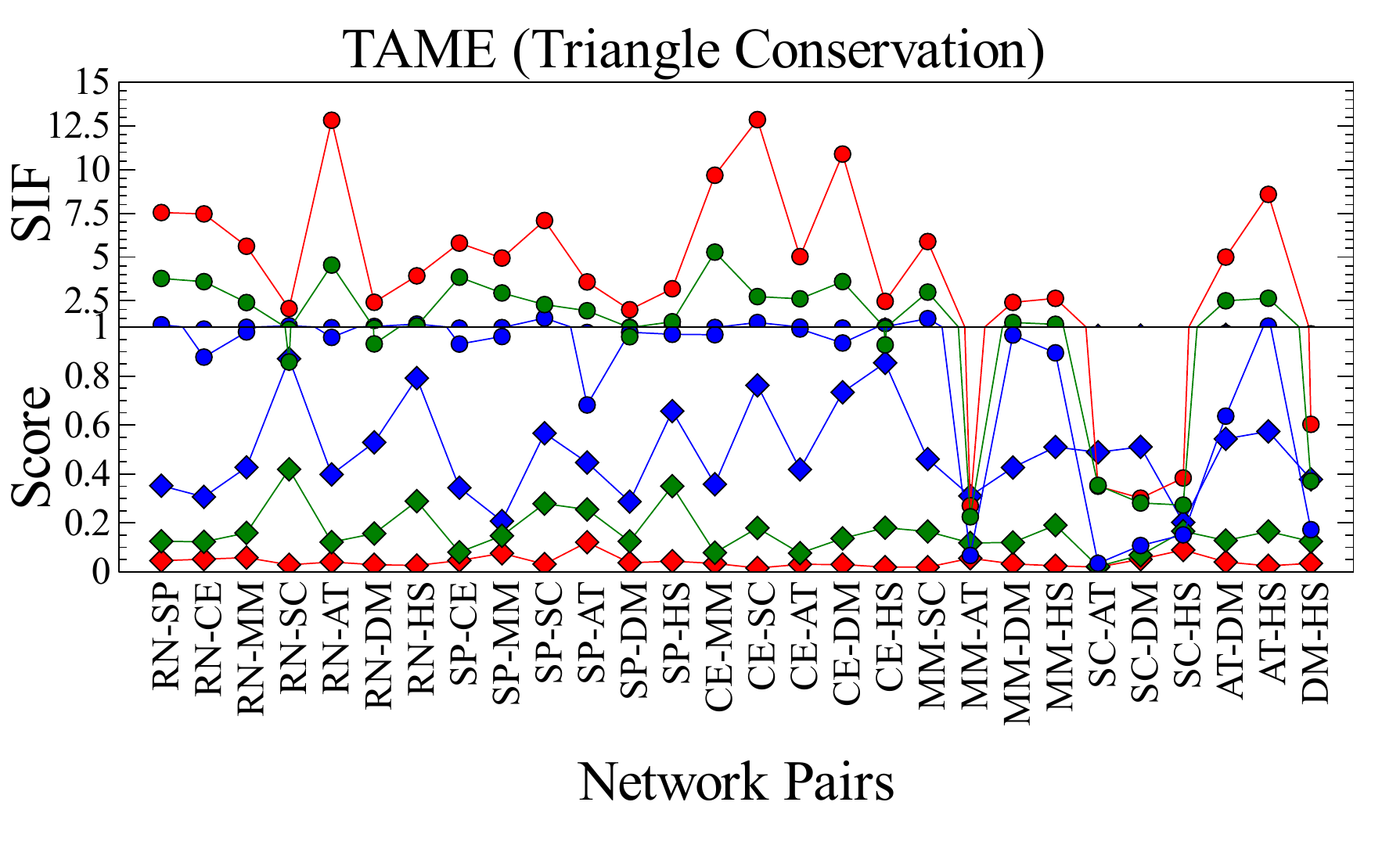}
    \caption{The chart of TAME's comparison against SANA. It follows the same pattern as the first two columns of figure 2 of the main publication.}
    \label{fig:tamegraphs}
\end{figure}

Figure \ref{fig:tamegraphs} includes the chart for SANA's comparison against only TAME. It follows the key in Figure 1 of the main publication.

\section{CPU Analysis}

The network alignment algorithms take a variable amount of time to align networks. Table \ref{tab:cpu} shows how long each network aligner took for each pair of networks. We performed all tests on AMD Opteron 6378 processors. We only count the time that each algorithm took to make the alignment; any overhead or preparation is not included. ModuleAlign is not included on this table because their code is too difficult to run in parallel properly and takes far too long to run in series than is worth it. We originally ran ModuleAlign in series but forgot to measure the exact time and rerunning would take too long than is important. ModuleAlign took around an hour to sometimes more than ten hours.

\begin{table*}
\caption{A complete runtime summary of all comparisons. The network aligners are organized from left to right by speed. All numbers are in seconds. Some network aligners, OptNetAlign and LGRAAL, used a user-specified amount of time.}
\label{tab:cpu}
\centering
\footnotesize
\begin{tabular}{| c c c c c c c c c c c |} 
 \hline
 
 Pair & PROPER & Hubalign & WAVE & SANA$^1$ & GHOST$^2$ & LGRAAL & OptNetAlign & NATALIE 2.0 & GREAT & MAGNA++ \\
\hline
RN-SP & 4.33 & 10.563 & 46.439 & 1200.982 & 218.272 & 3626.067 & 21627.105 & 36318.183 & 1782 & 6588.55 \\
\hline
RN-CE & 2.19 & 14.884 & 67.499 & 1205.878 & 254.168 & 3602.511 & 21594.192 & 36798.131 & 1710.79 & 7946.815 \\
\hline
RN-MM & 7.36 & 18.619 & 89.104 & 1204.306 & 357.098 & 3604.936 & 21597.395 & 37682.798 & 1584.07 & 10036.277 \\
\hline
RN-SC & 141.93 & 86.6 & 108.63 & 1207.698 & 697.682 & 3678.167 & 21620.571 &  &  & 37620.052 \\
\hline
RN-AT & 6.8 & 22.609 & 116.183 & 1202.232 & 494.11 & 3653.807 & 21598.76 & 39227.251 & 2464.4 & 12815.721 \\
\hline
RN-DM & 5.21 & 43.941 & 151.403 & 1202.941 & 761.318 & 3646.642 & 21594.732 & 6252.963 & 4954.68 & 23582.27 \\
\hline
RN-HS & 43.25 & 57.022 &  & 1204.202 & 1576.823 & 3718.981 & 21596.822 &  &  & 59503.783 \\
\hline
SP-CE & 3.08 & 18.746 & 86.451 & 1205.717 & 417.435 & 3618.937 & 21612.816 & 37658.221 & 5909.08 & 9233.363 \\
\hline
SP-MM & 4.35 & 23.241 & 118.53 & 1207.585 & 568.282 & 3621.64 & 21590.237 & 39494.238 & 5926.76 & 11278.529 \\
\hline
SP-SC & 32.42 & 54.946 & 142.788 & 1211.021 & 921.838 & 4035.292 & 21616.769 &  &  & 41411.814 \\
\hline
SP-AT & 5.71 & 32.811 & 147.993 & 1207.724 & 733.266 & 3790.599 & 21632.481 & 9208.076 & 7320.69 & 14183.744 \\
\hline
SP-DM & 7.8 & 57.022 & 206.362 & 1203.379 & 908.696 & 3757.407 & 21650.206 & 12863.58 & 11967.41 & 25254.116 \\
\hline
SP-HS & 31.28 & 95.469 &  & 1204.168 & 2020.892 & 3820.667 & 21644.772 &  &  & 61487.093 \\
\hline
CE-MM & 6.94 & 96.675 & 286.315 & 1208.865 & 934.614 & 3663.708 & 21641.083 & 5541.833 & 15797.56 & 12687.727 \\
\hline
CE-SC & 36.81 & 133.092 & 346.997 & 1211.463 & 1406.377 & 3647.845 & 21619.654 &  &  & 41783.25 \\
\hline
CE-AT & 7.46 & 96.456 & 370.909 & 1203.619 & 1203.285 & 3618.605 & 21620.149 & 6869.122 & 17449.32 & 15844.895 \\
\hline
CE-DM & 16.16 & 139.15 & 481.428 & 1201.269 & 1544.476 & 3720.279 & 21656.884 & 15684.91 & 19398.99 & 26565.248 \\
\hline
CE-HS & 49.9 & 235.182 &  & 1207.018 & 2979.279 & 3750.567 & 21581.82 &  &  & 62952.649 \\
\hline
MM-SC & 247.92 & 290.84 & 667.415 & 1203.642 & 2383.905 & 3890.048 & 21648.963 &  &  & 44150.335 \\
\hline
MM-AT & 22.02 & 269.954 & 770.317 & 1202.983 & 1888.38 & 3902.632 & 21632.147 & 10884.46 & 59376.92 & 18204.402 \\
\hline
MM-DM & 24.14 & 438.859 & 914.006 & 1206.956 & 2518.58 & 3687.144 & 21597.686 & 23804.774 & 66183.56 & 29844.522 \\
\hline
MM-HS & 110.46 & 621.394 &  & 1206.642 & 4655.939 & 5872.466 & 21618.983 &  &  & 65119.34 \\
\hline
SC-AT & 221.57 & 660.456 & 1131.38 & 1211.012 & 7490.71 & 4020.427 & 21760.065 &  &  & 48313.407 \\
\hline
SC-DM & 59.23 & 593.628 & 1495.722 & 1214.728 & 9495.281 & 3799.839 & 21689.732 &  &  & 62692.563 \\
\hline
SC-HS & 176.36 & 1080.103 &  & 1210.358 & 13686.569 & 5248.067 & 21773.541 &  &  & 100999.695 \\
\hline
AT-DM & 46.91 & 648.73 & 1642.546 & 1205.803 & 3653.613 & 3962.908 & 21602.428 & 142527.574 & 239526.32 & 33663.098 \\
\hline
AT-HS & 182.43 & 862.562 &  & 1216.563 & 6687.567 & 4718.554 & 21608.316 &  &  & 68265.577 \\
\hline
DM-HS & 96.58 & 1735.286 &  & 1223.887 & 12888.451 & 6127.944 & 21641.357 &  &  & 82430.435 \\
\hline
Average & 57.164 & 301.387 & 447.067 & 1207.237 & 2976.675 & 3993.096 & 21631.060 & 30721.074 & 30756.837 & 36944.974 \\
\hline
\end{tabular}
$^1$SANA takes a few extra minutes at the start of each run to calculate the temperature schedule. If this is included, SANA takes around 1400 seconds.

$^2$GHOST took almost 2 years of CPU time (11 days on 64 cores) in some cases to make the spectral signatures for individual networks. This time is not included in this chart.
\end{table*}


\bibliography{document}{}
\bibliographystyle{natbib}